\documentclass[sigconf,natbib=true,screen=true]{acmart}

\usepackage{color}
\usepackage{bbm}
\usepackage{multirow}
\usepackage[inline]{enumitem}
\usepackage{graphicx}
\usepackage{subcaption}
\usepackage{sistyle}
\usepackage[ruled]{algorithm2e}
\usepackage{booktabs}
\usepackage{caption}
\SIthousandsep{,}
\usepackage{makecell}
\usepackage{amsmath}
\usepackage[english]{babel}
\usepackage{hyperref}
\usepackage{array} 
\usepackage{graphicx}
\usepackage{algorithmic}
\usepackage{tikz}
\usepackage{wrapfig}
\usepackage{acronym}
\newcommand{\heading}[1]{\vspace*{.75mm}\noindent\textbf{#1.}}
\newcommand{\naming}[1]{\vspace*{.75mm}\noindent\textbf{#1}}

\AtBeginDocument{%
  \providecommand\BibTeX{{%
    \normalfont B\kern-0.5em{\scshape i\kern-0.25em b}\kern-0.8em\TeX}}}

\makeatletter
\g@addto@macro\normalsize{%
  \abovedisplayskip 3pt plus1pt 
  \belowdisplayskip 3pt plus1pt
  \abovedisplayshortskip  0pt plus1pt%
  \belowdisplayshortskip  0pt plus1pt
}
\makeatother

\acrodef{CV}{computer vision}
\acrodef{IR}{information retrieval}
\acrodef{LLM}{large language model}
\acrodef{MDP}{Markov decision process}
\acrodef{NLP}{natural language processing}
\acrodef{NRM}{neural ranking model}
\copyrightyear{2025}
\acmYear{2025}
\setcopyright{rightsretained}
\acmConference[SIGIR '25]{Proceedings of the 48th International ACM SIGIR Conference on Research and Development in Information Retrieval}{July 13--18, 2025}{Padua, Italy}
\acmBooktitle{Proceedings of the 48th International ACM SIGIR Conference on Research and Development in Information Retrieval (SIGIR '25), July 13--18, 2025, Padua, Italy}
\acmDOI{}
\acmISBN{}

\makeatother


\author{Yu-An Liu}
\orcid{0000-0002-9125-5097}
\affiliation{
	\institution{ICT, Chinese Academy of Sciences}
	\country{China}
}
\email{liuyuan21b@ict.ac.cn}

\author{Haya Nachimovsky}
\orcid{0009-0006-4383-7830}
\affiliation{
	\institution{Technion}
	\country{Israel}
}
\email{haya.nac@campus.technion.ac.il}

\author{Ruqing Zhang}
\orcid{0000-0003-4294-2541}
\affiliation{
	\institution{ICT, Chinese Academy of Sciences}
	\country{China}
}
\email{zhangruqing@ict.ac.cn}

\author{Oren Kurland}
\orcid{0000-0002-0669-0431}
\affiliation{
	\institution{Technion}
	\country{Israel}
}
\email{kurland@technion.ac.il}

\author{Jiafeng Guo}
\orcid{0000-0002-9509-8674}
\affiliation{
	\institution{ICT, Chinese Academy of Sciences}
	\country{China}
}
\email{guojiafeng@ict.ac.cn}

\author{Moshe Tennenholtz}
\orcid{0000-0002-9459-5388}
\affiliation{
	\institution{Technion}
	\country{Israel}
}
\email{moshet@technion.ac.il}

\begin{document}

\title[Robust-IR @ SIGIR 2025: The First Workshop on Robust Information Retrieval]{Robust-IR @ SIGIR 2025: The First Workshop on \\ Robust Information Retrieval}

\begin{abstract}
With the advancement of information retrieval (IR) technologies, robustness is increasingly attracting attention. 
When deploying technology into practice, we consider not only its average performance under normal conditions but, more importantly, its ability to maintain functionality across a variety of exceptional situations. 
In recent years, the research on IR robustness covers theory, evaluation, methodology, and application, and all of them show a growing trend. 
The purpose of this workshop is to systematize the latest results of each research aspect, to foster comprehensive communication within this niche domain while also bridging robust IR research with the broader community, and to promote further future development of robust IR. 
To avoid the one-sided talk of mini-conferences, this workshop adopts a highly interactive format, including round-table and panel discussion sessions, to encourage active participation and meaningful exchange among attendees.

\end{abstract}

\maketitle

\section{Title}
Robust-IR @ SIGIR 2025: The First Workshop on Robust Information Retrieval\footnote{\url{https://sigir-2025-workshop-on-robust-ir.github.io/}}

\section{Motivation}
Information retrieval (IR) systems serve as a fundamental tool for enabling users to access relevant information efficiently.
In recent years, the rise of deep learning has revolutionized IR systems, with deep neural networks demonstrating impressive effectiveness in various tasks \cite{guo2016deep,liu2017cascade,zhao2022dense,guo2022semantic}. 
However, alongside these advancements, neural IR models also reveal serious vulnerability flaws \cite{wu2022prada,thakur2021beir,cohen2018cross}, which significantly limit their reliability and scalability in real-world applications.

IR systems are typical scenarios exposed to robustness challenges.
For instance, search engines are natural scenarios for competitive search, where malicious ranking attacks driven by search engine optimization (SEO) tactics are prevalent \cite{gyongyi2005web, kurland2022competitive,nachimovsky2024ranking}. 
Additionally, IR systems need to continuously adapt to dynamic scenarios, such as updates to the corpus and the influx of unseen queries, which introduce out-of-distribution (OOD) data challenges \cite{thakur2021beir, wu2022neural,cohen2018cross,yu2022coco}.
The robustness of IR systems has therefore become an increasingly prominent research focus. 
A range of strategies and methods have been proposed to mitigate these issues \cite{chen2022towards,wu2022prada,liu2022order,liu2023topic}, laying the groundwork for more robust and dependable IR systems.

Despite these advancements, research on IR robustness remains fragmented.
Research fields including theory, evaluation, methodology, application, and society are developed separately and are not referenced to each other due to their different focuses.
This fragmentation limits the ability to leverage advancements across domains effectively.
Given that robust IR has now reached a significant level of maturity, fostering open communication and knowledge sharing is crucial. 
Members of our team have previously organized the robust IR tutorial \cite{liu2024robust} to introduce this field to the broad community. 
Given the lack of dedicated workshops in this field, we are eager to seize this opportunity to foster deeper exchanges.

This workshop aims to serve as a bridge for researchers across various fields of robustness to engage in meaningful and productive exchange, while simultaneously providing new perspectives and inspiration for IR research.
By fostering a virtuous cycle, i.e., \emph{theoretical insights guide method design, practical methods find application in real-world scenarios, and real-world applications inspire further theoretical advancements}, we hope to drive forward the development of robust IR systems. 
Ultimately, we seek to encourage a broader community of researchers to join the effort in advancing robustness in IR, paving the way for sustained innovation and growth in this critical area.
\section{Theme and Purpose of the Workshop}
The RobustIR workshop will discuss the robustness challenges that IR systems currently face or may encounter, the underlying theoretical mechanisms, and approaches to building robust and trustworthy IR systems.
The aim of this workshop is multifaceted:
\begin{enumerate}[label={(\arabic*)}, leftmargin=*]
\item To gather IR researchers and practitioners in gaining a deeper understanding of the robustness challenges of researching, developing, and deploying IR models in practice.
\item To provide a platform for exchanging and sharing the latest perspectives and research achievements in robust IR.
\item To identify key issues in robust IR and Inspire future directions.
\end{enumerate}
To build a systematic understanding of robust IR, we have structured the workshop agenda around the following five pillars below (Theory, Evaluation, Method, Application, and Society Impact).
We will provide more details for each theme in the following sections.

\subsection{Theory}
Design of theoretical frameworks to identify vulnerabilities in IR systems and propose novel solutions with theoretical guarantees.

\begin{enumerate}[leftmargin=*, label=\textbullet]
    \item \heading{Game theory} 
    Analysis of robustness from a game theory perspective, particularly in adversarial settings where attacks originate from strategic agents that exploit the vulnerabilities of the system. This research examines the incentives behind adversarial attacks and develops mechanisms to counteract them. Topics include modeling strategic interactions as games, designing mechanisms to mitigate adversarial behaviors, and understanding the implications of these strategies for robust system design.
    
    \item \heading{Competitive search}
    Investigating robustness in competitive environments, such as those between publishers, search engines, or advertisers. 
    This includes analyzing how competition influences the ecosystem and exploring mechanisms to promote desired properties in these contexts such as robustness and fairness. 

    \item \heading{Probability ranking principle}
    Revisiting the Probability Ranking Principle (PRP) in the context of robust information retrieval. This includes investigating the assumptions of PRP and their adaptation to different adversarial scenarios.

\end{enumerate}

\subsection{Evaluation method}
Evaluation is important to understand the robustness of a model.
Robustness is a general topic that often covers multiple aspects. 
\begin{enumerate}[leftmargin=*, label=\textbullet]
    \item \heading{Specific evaluation}
    Robustness evaluation for specific robustness types (e.g., adversarial, OOD robustness), scenarios (e.g., corpus updating, queries with typos), and IR models (e.g., sparse, dense, and generative retrieval models).
    \item \heading{General evaluation}
    Using an evaluation metric to comprehensively cover as many robustness scenarios as possible.
    \item \heading{Diverse evaluation tools}
    Developing new evaluation forms, including new evaluation functions, LLMs, and human evaluations, for comprehensive robustness comparisons.
    \item \heading{Benchmarks of robustness}
    Discussion of existing robustness datasets and proposing new benchmark tasks to address diverse robustness categories and requirements.
\end{enumerate}

\subsection{Method}
Developing effective methods to improve the robustness of IR systems across different settings is an important issue.

\begin{enumerate}[leftmargin=*, label=\textbullet]
    \item \heading{Adversarial attack \& defense}
    Investigating adversarial vulnerabilities in IR models and developing defenses against malicious attacks like data poisoning and adversarial document generation.
    \item \heading{Zero/few shot IR}
    Enhancing OOD robustness in low-resource settings with scarce labeled data. This involves using zero-shot or few-shot learning, transfer learning, and large-scale pre-trained models to improve cross-domain and task generalization.
    \item \heading{Balancing robustness and effectiveness}
    Trade-offs exist between robustness and effectiveness \cite{tsipras2019robustness,rozsa2016accuracy}. We encourage exploring enhancing robustness without compromising effectiveness.
    \item \heading{Long-term learning}
    IR systems must adapt to dynamic environments with new topics and unknown queries. This subtopic explores continual learning to enhance stability.
    \item \heading{Noise Resistance}
    Making IR systems resistant to noise in queries, documents, or training data, such as typo handling, semantic noise filtering, and processing incomplete or corrupted inputs.
    \item \heading{Enhancing RAG Robustness}
    RAG systems combine retrieval and generative components. This subtopic aims to improve RAG pipeline robustness by reducing error propagation, ensuring consistency between retrieved documents and generated content, and enhancing output reliability under uncertain conditions.
\end{enumerate}

\subsection{Application}
The robustness issue in practical applications and the applicability of robustness enhancement methods in reality.

\begin{enumerate}[leftmargin=*, label=\textbullet]
    \item \heading{Robust search engines}
    Deploying robustness enhancement methods in resource- and condition-constrained search engines.
    \item \heading{Robust recommendation systems}
    Robustness for recommendation systems, addressing sparse user data, cold-start issues, adversarial manipulation, and dynamic user preferences.
    \item \heading{Data-specific scenarios}
    Robustness in specialized data retrieval, including scientific literature, medical documents, legal data, and long-form or multi-modal documents.
    \item \heading{Federated and distributed IR systems}
    Robustness for distributed IR using federated learning involves tackling inconsistent local data, implementing privacy-preserving strategies, and improving communication efficiency across distributed nodes.
\end{enumerate}

\subsection{Society Impact}
The societal impact of robustness in IR systems is far-reaching and requires careful consideration.

\begin{enumerate}[leftmargin=*, label=\textbullet]
    \item \heading{Human behaviors that affect robustness}
    Understand how user behaviors like query biases, click feedback loops, and manipulated web content affect IR system robustness.
    \item \heading{Robustness and Ethics}
    Addressing ethical concerns by ensuring fairness, minimizing algorithmic biases, and maintaining transparency. Ensuring robust models do not disproportionately affect certain user groups or perpetuate societal inequalities.
    \item \heading{Explainability and truthfulness}
    Discuss the impact of IR model explainability and information truthfulness on users.
\end{enumerate}
\section{Format and Planned Interaction}
RobustIR will be an interactive full-day hybrid workshop that avoids the one-sided dialogue of a mini-conference.
\begin{enumerate}[leftmargin=*, label=\textbullet]
    \item Invited keynote talk (industrial and academic) [hybrid].
    \item Invited panel discussions (industrial and academic) [hybrid].
    \item Hybrid paper presentation session [hybrid] for accepted papers.
    \item A round-table discussion to share lessons learned [onsite].
\end{enumerate}

\begin{table}[h]
\centering
\caption{Timetable}
\begin{tabular}{c l}
\toprule
\textbf{Time} & \textbf{Activity} \\
\midrule
Morning & \\
\midrule
09:00 – 09:30 & Opening \\
09:30 – 10:30 & Keynote \\
10:30 – 11:00 & Coffee break \\
11:00 – 12:00 & Keynote \\
12:00 – 12:30 & Paper Presentation \\
12:30 – 13:30 & Lunch \\
\midrule
Afternoon & \\
\midrule
13:30 – 14:30 & Paper Presentation \\
14:30 – 15:30 & Panel Discussions \\
15:30 – 16:00 & Coffee break \\
16:00 – 16:45 & Roundtable Discussion \\
16:45 – 17:00 & Closing \\
\bottomrule
\end{tabular}
\end{table}

\section{Distinction from Main Conference Topics}
The core algorithm of IR, search and ranking, is one of the topics of this main conference. 
Robust IR focuses on the robustness of search and ranking algorithms from both theoretical and empirical perspectives, benefiting web search and complementing the understanding of search and ranking.
Machine learning for IR, including deep learning and generative models, is another topic of the main conference.
Current research generally focuses on their effectiveness and efficiency. 
Robustness offers a broader perspective and complements the conference's focus on fairness, accountability, transparency, ethics, and explainability (FATE).
\section{Organizers}
ICT and Techniuon teams, respectively, have explored in depth the methodological and evaluation, theory and society impact of robust IR. 
Their combined expertise can help advance this workshop.

\naming{Yu-An Liu} is a Ph.D. student at the Institute of Computing Technology, Chinese Academy of Sciences. 
His research centers on the adversarial and OOD robustness of IR systems.
He has conducted robust IR tutorials\footnote{\url{https://sigir2024-robust-information-retrieval.github.io/}} at SIGIR'24 \cite{liu2024robust} and will organize another at WSDM'25 \cite{liu2025robust}, again as a presenter with Ruqing Zhang and Jiafeng Guo.
He is the first author of several full papers on adversarial robustness \cite{liu2023topic, liu2023black, liu2024perturbation, liu2024multi, liu2025attachain}, OOD robustness \cite{liu2023robustness, liu2025robustness} in IR, as well as a survey paper \cite{liu2024robust_survey} on robust IR.

\naming{Haya Nachimovsky} is a Ph.D. student at the Technion. Her research focuses on information retrieval in competitive environments, combining game-theoretic analysis with empirical studies to understand strategic behaviors and propose practical solutions \cite{nachimovsky2024corpusenrichment, nachimovsky2024ranking,nachimovsky2024competitive,nachimovsky2025multi,mordo2025csp}.
She received the ICTIR'24 Best Paper Award for her work providing theoretical and empirical insights into how document publishers adapt to rankings in competitive settings.

\naming{Ruqing Zhang} is an Associate Researcher at the Institute of Computing Technology, Chinese Academy of Sciences. 
Her recent research focuses on generative retrieval and robustness in the context of IR. 
She has received the EMNLP'24 Best Paper Award about the safety of LLMs and authored several papers in the field of robust IR  \cite{liu2023topic, liu2023black, liu2024perturbation, liu2024multi, liu2025attachain,liu2023robustness, liu2025robustness,liu2024robust_survey,wu2022neural,wu2022prada,wu2022certified}. 
She was a co-organizer of tutorials and workshops at SIGIR, Webconf, ECIR, and SIGIR-AP, e.g., Gen-IR workshops at SIGIR'23 and SIGIR'24 \cite{benedict2023gen,benedict2024gen}, Gen-IR tutorials at SIGIR-AP'23/Webconf'24/ECIR'24/SIGIR'24 and Robust-IR tutorials at SIGIR'24 and WSDM'25.

\naming{Oren Kurland} is a professor at the Technion. He is a distinguished member of the ACM. Throughout the last few years, Oren has worked on competitive search; specifically, integrating game theoretic models in information retrieval. Oren served as a program co-chair for ICTIR and for the short papers track in SIGIR. He was awarded a best paper award in ICTIR and a runner-up award in SIGIR and ECIR. Oren served as an area chair/senior program committee member for SIGIR, WSDM, the Web Conf, CIKM, and ECIR. He also served on the editorial board of the Information Retrieval Journal.

\naming{Jiafeng Guo} is a Researcher at the Institute of Computing Technology, Chinese Academy of Sciences, and a Professor at the UCAS. 
He is the director of the CAS key lab of network data science and technology. 
He has worked on a number of topics related to Web search and data mining, with a current focus on neural models for IR and NLP. 
He has received multiple best paper (runner-up) awards at leading conferences (CIKM’11, SIGIR’12, CIKM’17, WSDM’22, EMNLP'24). 
He has been (co)chair for many conferences, e.g.,  reproducibility track co-chair of SIGIR'23, workshop co-chair of SIGIR'21, and short paper co-chair of SIGIR'20. 
He served as an associate editor for ACM Transactions on Information Systems and Information Retrieval Journal. 
Jiafeng has previously taught tutorials at many IR-related conferences.

\naming{Moshe Tennenholtz} is a professor at the Technion specializing in the intersection of AI, game theory, and economics. 
Moshe has founded the Microsoft Research activity in Israel, the Technion-Microsoft research center, and served as chief scientist of AI21 lab. 
Moshe served as Editor-in-Chief of JAIR and associate editor for Games and Economic Behavior, Artificial Intelligence, the international journal of autonomous agents and multi-agent systems, and the ACM Transactions on Economics and Computation. 
Moshe is a AAAI fellow, an ACM fellow, a fellow of the society for advancement of economic theory, and a council member of the game theory society.
He is a winner of the ACM Allen Newell award, a winner of the John McCarthy award, and a winner of the ACM/SIGART Autonomous Agents Award. 
In joint work, he introduced contributions to the interplay between artificial intelligence and game theory/economics, like studies of artificial social systems, co-learning, etc., and the first near-optimal algorithm for adversarial reinforcement learning.
Moshe's recent work on mechanism design for data science earned him an ERC Advanced Grant.
\section{PC Members}
Potential PC members for reviewing paper submissions:
\begin{itemize}
    \item Shengyao Zhuang, CSIRO
    \item Ben He, University of Chinese Academy of Sciences
    \item Zexuan Zhong, Princeton University
    \item Andrew Parry, University of Glasgow
    \item Jinyan Su, Cornell University
    \item Wei Zou, Pennsylvania State University
    \item Runpeng Geng, Pennsylvania State University
    \item Jiawei Liu, Wuhan University
\end{itemize}


\section{Selection Process}
We invite submissions of research, position, demo, and opinion papers through an open call. Papers should be either short (up to 4 pages) or full-length (up to 9 pages). 
The program committee will peer-review all submissions for relevance to the workshop and potential to spark discussion.

\section{Audience}
\heading{Audience reach}
The robustness of IR models has been of high interest to industrial practitioners, who may join the workshop to discuss challenges and solutions. 
Meanwhile, robust IR is gaining traction in academia, drawing novice researchers eager for discussion. 
Robustness is also a prominent topic in fields like computer vision (CV) and natural language processing (NLP), potentially attracting scholars interested in cross-disciplinary collaboration.

\heading{Target Audience}
The target audience is academia/industry researchers working on or interested in robust information retrieval. 
We will promote the workshop on social media such as X and prepare a dedicated website for the last edition.

\section{Related Workshops}
Related workshops include:
\begin{enumerate}[label=(\roman*)]
    \item Reproducibility, Inexplicability, and Generalizability of IR \\
          \textnormal{[RIGOR Workshop @ SIGIR 2015]}
    \item Explainable Recommendation and Search \\
          \textnormal{[EARS Workshop @ SIGIR 2018]}
    \item Privacy, Security and Trust in Computational Intelligence  \\
          \textnormal{[PSTCI Workshop @ CIKM 2021]}
    \item LLMs' Interpretability and Trustworthiness \\
          \textnormal{[LLMIT Workshop @ CIKM 2023]}
\end{enumerate}
Our workshop focuses on robust IR, with specific emphasis on its theories, methodologies, evaluation, applications, and society.

\bibliographystyle{ACM-Reference-Format}
\balance
\bibliography{references}

\end{document}